\documentclass[aps,prl,twocolumn,epsfig,groupedaddress]{revtex4}

\usepackage{epsfig}

\def\beq{\begin{equation}}
\def\eeq{\end{equation}}
\def\bea{\begin{eqnarray}}
\def\eea{\end{eqnarray}}

\def\eq#1{{Eq.~(\ref{#1})}}
\def\fig#1{{Fig.~\ref{#1}}}
\def\ud{\underline}

\def\iq{\!=\!}
\begin{document}

\title{Particle multiplicities in Lead-Lead collisions at the
  LHC from non-linear evolution with running coupling}

\author{Javier L. Albacete}
\address{
Department of Physics, The Ohio State University,
191 W. Woodruff Avenue, OH 43210, Columbus, USA}
\date{\today}

\begin{abstract}
We present predictions for the pseudo-rapidity density of charged particles
produced in central Pb-Pb collisions at the LHC. Particle production
in such collisions is calculated in the 
framework of $k_t$-factorization. The nuclear unintegrated
gluon distributions at LHC energies are determined from numerical
solutions of the Balitsky-Kovchegov equation including recently
calculated running coupling corrections. The initial 
conditions for the evolution are fixed by fitting RHIC data at
collision energies $\sqrt{s_{NN}}=130$ and $200$ GeV per nucleon. We
obtain $dN^{Pb-Pb}_{ch}/d\eta\,(\sqrt{s_{NN}}=5.5\,
\mbox{TeV})\vert_{\eta=0}\approx 1290\div 1480$.
\end{abstract}
\maketitle
It has been suggested that the nucleus-nucleus collisions
performed at the Relativistic Heavy Ion Collider (RHIC) at the highest
collision energies of  $130$ and 200 GeV per nucleon probe the Color
Glass Condensate \cite{cgc} regime of QCD governed by non-linear 
coherent phenomena and gluon saturation. This claim is
supported by the success 
of saturation models \cite{klm,kl,kln,asw,hn} in the description of the
energy, rapidity 
and centrality dependence of the particle multiplicities
experimentally measured in d-Au and Au-Au collisions. 
With collision energies of up to $5.5$ TeV, the upcoming program in
lead-lead collisions at the CERN Large Hadron Collider (LHC) is expected to
provide confirmation for the tentative  
conclusions reached at RHIC and to discriminate between the different
physical mechanisms proposed to explain particle production in high
energy nuclear reactions (for a review of alternative approaches see,
 e.g., \cite{ap}). 

The phenomenological models in \cite{klm,kl,kln,asw} rely on the
assumption that 
the saturation  scale $Q_{sA}$ that governs the onset of non-linear
effects in the wave function 
of the colliding nuclei is perturbatively large $\sim\!1\!$ GeV at the highest
RHIC energies. Next, gluon production is calculated via the
convolution of the nuclear unintegrated gluon distributions (ugd's)
according to $k_t$-factorization \cite{glr}. Under the additional assumption of
local parton-hadron 
duality, the multiplicity
in A-A collisions at central rapidity rises proportional to the
saturation scale,
$dN^{AA}/d\eta\vert_{\eta=0}\propto Q^2_{sA}$, \cite{yuri}. On the other hand, the
growth of the saturation scale with increasing energy (equivalently,
decreasing Bjorken-$x$) is determined by the
perturbative BK-JIMWLK non-linear evolution equations \cite{bk,jimwlk} (for a
complete set of references see \cite{cgc}), thereby
establishing a direct link between the 
initial state gluon saturation dynamics and the 
experimentally measured hadron yields. 
The energy dependence of the saturation scale yielded by the BK-JIMWLK
equations at the degree of 
accuracy of their original derivation, leading-logarithmic (LL) in
$\alpha_s\ln(1/x)$ with $\alpha_s$ fixed, is $Q_s^2\approx
Q_0^2(x_0/x)^{\lambda}$ with $\lambda\approx 4.8\frac{N_c}{\pi}\alpha_s$
\cite{aamsw,i^2m}. This growth is too fast to be 
reconciled with the energy dependence observed in RHIC multiplicity data, which
indicate $\lambda\!\sim\!0.2\div0.3$ \cite{hn,kln,asw,rhic}. 
Such deficiency of the theory has been circumvented so far by 
leaving $\lambda$ as a
free parameter, often adjusted to the empirical value
$\lambda\!\approx 
\!0.288$ obtained in fits to small-$x$ HERA data in deeply inelastic
lepton-proton scattering in the framework of saturation models \cite{gbw,i^2m}.

Higher order corrections to the BK-JIMWLK equations have
been calculated recently via all orders resummation of
$\alpha_sN_f$ contributions \cite{bkrc}. Such corrections bring
substantial modifications to the LL kernel, including
running coupling effects, and result in a significant slowdown in the
speed of the evolution, among other quantitatively important dynamical
effects \cite{Albacete:2007yr}. 

In this work we demonstrate that this partial improvement
is sufficient to describe the 
energy and rapidity dependence of the multiplicities in Au-Au
collisions at the highest RHIC energies. Then we extrapolate to
LHC energies and present predictions for Pb-Pb collisions at
$\sqrt{s_{NN}}=5.5$ TeV.

We start by solving the non-linear small-$x$ evolution equation
for the dipole-nucleus scattering 
matrix, $S(Y,r)$, including running coupling corrections
\cite{bkrc,Albacete:2007yr}:   
\begin{equation}\label{frs}
  \frac{\partial S(Y,r)}{\partial Y} \, = \,
  \mathcal{R}\left[S\right]-\mathcal{S}\left[S\right]\,,
\end{equation}
where $r$ is the dipole size and $Y\iq\ln(x_{0}/x)$. The first, {\it
  running coupling}, term of the evolution kernel, $\mathcal{R}[S]$, recasts the
higher order corrections that amount to a modification of the LL
small-$x$ gluon emission kernel, leaving the interaction structure of
the LL equation untouched, whereas the second, {\it 
  subtraction}, term, $\mathcal{S}[S]$, 
accounts for the new interaction 
channels opened up by the higher order corrections. Explicit
expressions for both terms as well as a detailed explanation of the
numerical method used to solve \eq{frs} are given in
\cite{Albacete:2007yr}.     
The initial conditions for the evolution are taken from the
semi-classical McLerran-Venugopalan (MV) model \cite{mv}, aimed at
describing the gluon distributions of 
large nuclei at moderate values of Bjorken-$x$, prior to the onset of
quantum corrections. Thus, the initial dipole scattering
amplitude, $\mathcal{N}\iq 1\!-\!S$, reads    
\begin{equation}
\mathcal{N}(Y\!=\!0,r)=
1-\exp\left\{-\frac{r^2Q_0^2}{4}\ln\left(\frac{1}{r\Lambda}+e\right)\right\},  
\label{nmv}
\end{equation}
where $Q_0$ is the initial saturation scale. The constant $e$ under the
logarithm acts as an infrared regulator and $\Lambda\iq0.2$ GeV. 

\begin{figure}[h]\epsfxsize=8.7cm
\centerline{\epsfbox{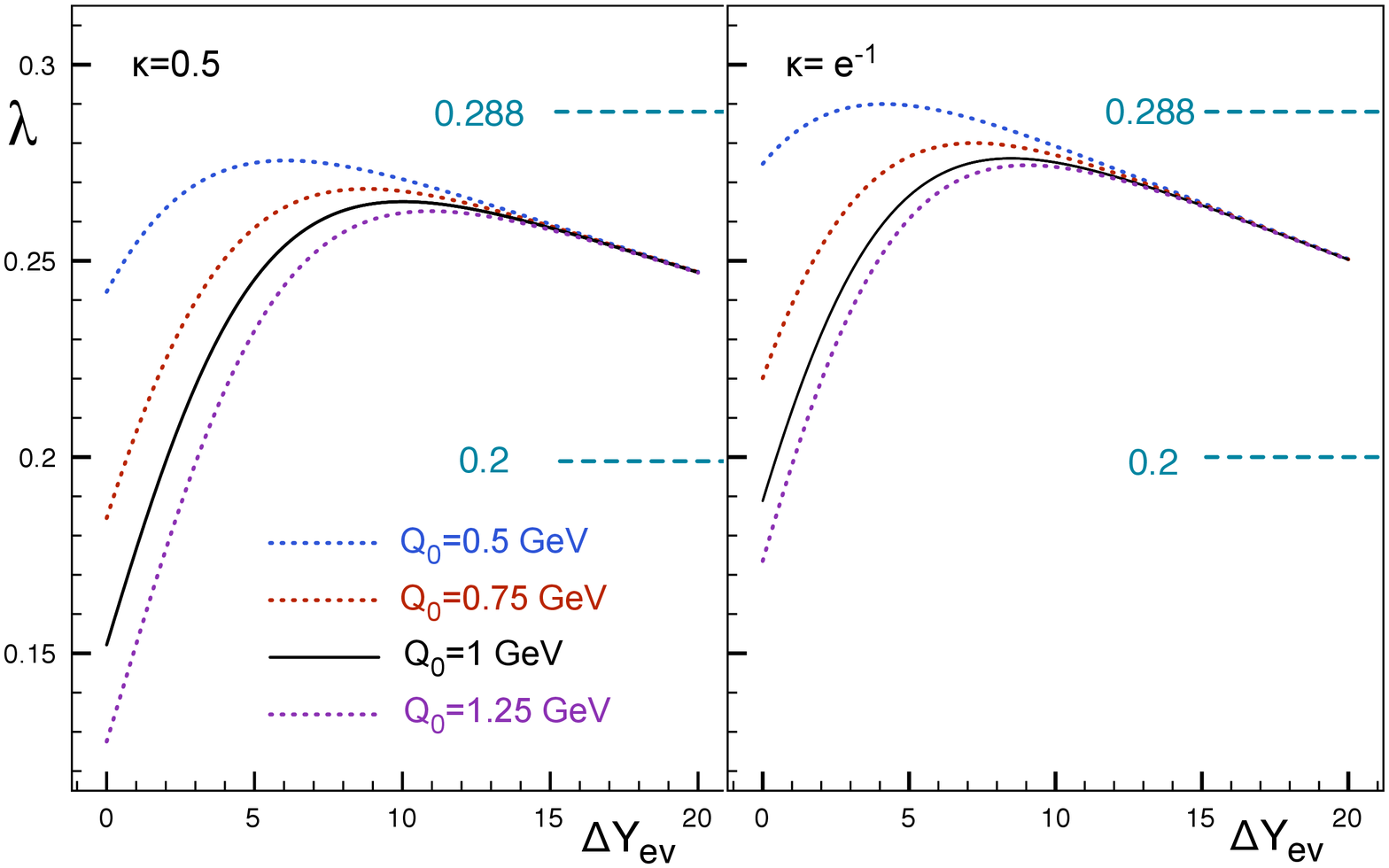}}
\caption{
  $\lambda\iq\frac{d\ln Q_s^2(Y)}{dY}$, for $Q_{0}\iq0.5$, 0.75, 1 and
  1.25 GeV (from top to bottom), and for $\kappa\iq0.5$ (left) and
  $\kappa\iq e^{-1}$ (right).
}\label{fig1}
\end{figure}

The {\it speed} of evolution, $\lambda\iq d\ln Q^2_s(Y)/dY$, extracted from
numerical solutions of \eq{frs} corresponding to different initial
conditions ($Q_0\iq0.5$, 0.75, 1 and 1.25 GeV) is plotted in \fig{fig1}. 
For $Y\!>\!0$ the saturation scale is determined by the condition
$\mathcal{N}(Y,r\iq1/Q_s(Y))\iq\kappa$, with $\kappa\iq0.5$
(left plot) and $\kappa\iq e^{-1}$ (right plot).     
These results show two remarkable features of the solutions. 

First, the running coupling  corrections render the
energy dependence of the saturation scale compatible with the one
indicated by the analysis of experimental data.  
Thus, the $\lambda$ values in \fig{fig1} are slightly smaller 
than the one extracted from fits to HERA data, $\lambda\iq0.288$
(except, perhaps, for $Q_0\lesssim 0.5$ GeV at small rapidities). On
average, they are compatible with $\lambda\iq0.2$ reported in
\cite{hn} as the optimal 
value to reproduce the energy and rapidity dependence of the
multiplicities in Au-Au collisions at the highest RHIC energies.
Second, they reveal the existence of two very distinct kinematical
regimes: At small {\it pre-asymptotic} rapidities the evolution is
strongly dependent on the initial
conditions. In particular, denser systems, i.e. those associated to
larger values of $Q_0$, evolve more slowly due to
the relative enhancement of non-linear effects with respect to more
dilute systems. Such dependence on the nature of
the evolved system is completely washed out by the evolution and, at
high enough rapidities, all the solutions reach a common speed of
evolution. 
The onset of this universal {\it 
  scaling} regime is reflected in \fig{fig1} by the 
convergence of all the individual trajectories into a single
curve for $Y\gtrsim15$. The studies of more exclusive properties of
the solutions carried out in
\cite{aamsw, Albacete:2007yr} suggest that the full scaling
regime is reached at even larger rapidities,
$Y\!\gtrsim\!80$. Moreover, sizable
scaling violations have been detected in HERA data \cite{iim} and
in particle spectra in d-Au collisions at RHIC \cite{dhj}. These
observations rise the 
question of whether the scaling ansatz that connects HERA and RHIC
phenomenology through the universality property of the solutions is an
adequate one at presently available energies. 

\begin{figure}[h]\epsfxsize9.5cm
\centerline{\epsfbox{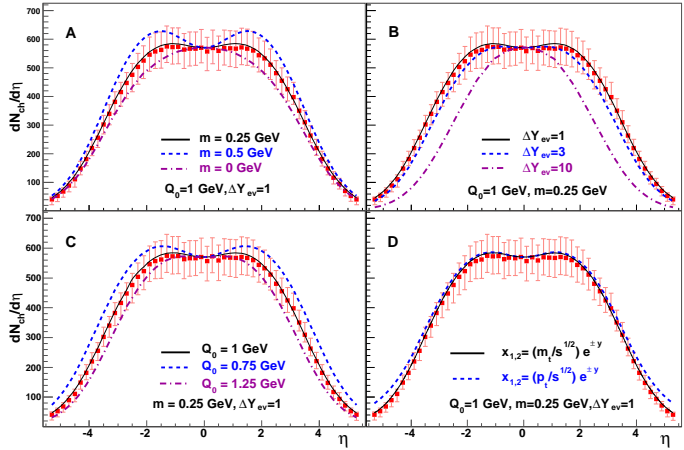}}
\caption{Pseudo-rapidity density of charged particles produced in Au-Au 0-6\%
  central collisions at $\sqrt{s_{NN}}=130$ GeV. Data taken from
  \cite{rhic}. The solid lines correspond 
  to $Q_{0}\!=\!1$
  GeV, $m\!=\!0.25$ GeV, $\Delta Y\!=\!1$ and
  $x_{1,2}\!=\!(m_t/\sqrt{s})\,e^{\pm y}$. The modifications to this
  central value considered are:
  {\it Upper-left}: $m\!=\!0.5$ GeV (dashed line) and
  $m\!=\!0$ GeV (dashed-dotted line).
  {\it Upper-right}:  $\Delta Y\!=\!3$ (dashed line) and $\Delta Y\!=\!10$
(dashed-dotted line).
 {\it Lower-left}: $Q_{0}\!=\!0.7$
  GeV, (dashed line) and $Q_{0}\!=\!1.25$ GeV, (dashed-dotted line).
 {\it Lower-right}: $x_{1,2}\!=\!(p_t/\sqrt{s})\,e^{\pm y}$ (dashed
  line).
} \label{fig2}
\end{figure}

In analogy to \cite{kln,asw}, we calculate the pseudo-rapidity
density of charged particles produced in nucleus-nucleus collisions
within the $k_t$-factorization framework via:  
\begin{eqnarray}
\frac{dN_{ch}}{dy\, d^2b}&=&C\frac{4\pi
  N_c}{N_c^2-1}\int\frac{d^2p_t}{p_t^2}\int^{p_t}\,d^2k_t\,
\alpha_s(Q)\nonumber \\
& &\times\,\varphi\left(x_1,\frac{\vert\ud{k_t}+\ud{p_t}\vert}{2}\right)
\varphi\left(x_2,\frac{\vert\ud{k_t}-\ud{p_t}\vert}{2}\right), 
\label{ktfact}
\end{eqnarray}
where $p_t$ and $y$ are the transverse momentum and rapidity of the
produced particle, $x_{1,2}\iq(p_t/\sqrt{s})\,e^{\pm y}$, 
$Q\iq0.5\max\left\{\vert p_t\pm k_t\vert \right\}$ and $b$ the impact
parameter of the collision. The lack of
impact parameter integration in this calculation and the gluon to
charged hadron ratio are accounted for by the constant $C$, which sets the
normalization. 
The nuclear unintegrated gluon distribution entering
\eq{ktfact} is related to the inclusive gluon distribution,
$\varphi(x,k)\propto\frac{d(xG(x,k^2))}{d^2k\,d^2b}$, 
and is given in terms of the dipole scattering amplitude evolved
according to \eq{frs}: 
\begin{equation}
   \varphi(Y,k) 
   = \int {d^2r\over 2\pi\, r^2}\exp\{i\, {\ud r}\cdot {\ud
     k}\}\,\mathcal{N}(Y,r)\, , 
   \label{phi}
\end{equation}
The relation between the evolution variable in \eq{frs} and
Feynman-$x$ of the produced particle is taken to be 
$Y\!=\!\ln(0.05/x_{1,2})+\Delta Y_{ev}$.
Since the relevant values of Bjorken-$x$ 
probed at mid-rapidities and $\sqrt{s_{NN}}\iq130$ GeV at RHIC
are estimated to be $\sim0.1\div0.01$, the free parameter $\Delta Y_{ev}$
controls the extent of evolution undergone by the nuclear gluon
densities resulting of \eq{frs} prior to comparison with RHIC data. 
Similar to \cite{kln}, large-$x$ effects have been modelled by replacing
$\varphi(x,k)\rightarrow \varphi(x,k)(1-x)^4$.
The running of the strong coupling, evaluated according to the one
loop QCD expression, is regularized in the infrared by freezing it to a
constant value $\alpha_{fr}\iq1$ at small momenta.
Finally, in order to compare \eq{ktfact} with
experimental data it is necessary to correct the difference between
rapidity, $y$, and the experimentally measured pseudo-rapidity,
$\eta$. This is achieved by introducing an average hadron mass,
$m$. The variable transformation, $y(\eta,p_t,m)$, and its
corresponding Jacobian are given by Eqs.(25-26) in 
\cite{kl}. Corrections to the kinematics due to
the hadron mass are also considered by replacing $p_t^2\rightarrow
m_t^2\iq p_t^2+m^2$ in the evaluation of $x_{1,2}$. Remarkably, the
optimal value found in comparison with data, $m\sim0.25$ GeV, see
\fig{fig2}A, is in good quantitative agreement with the 
hadrochemical composition of particle production at RHIC. 

\begin{figure}[h]\epsfxsize=8.7cm
\centerline{\epsfbox{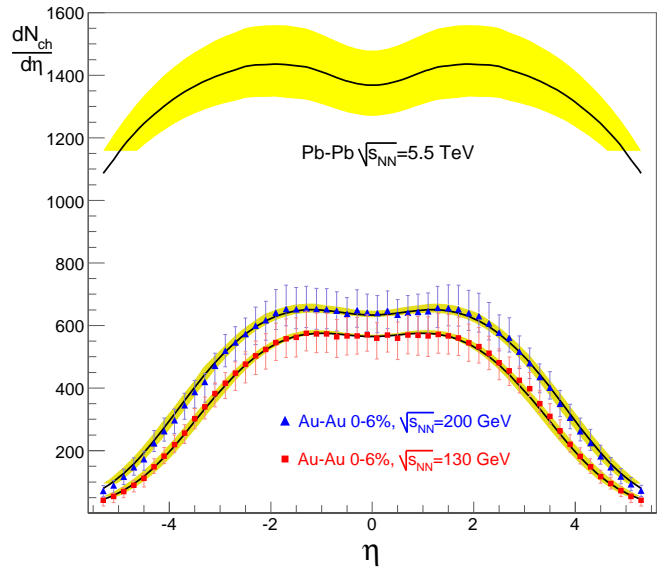}}
\caption{Pseudo-rapidity density of charged particles produced in Au-Au 0-6\%
  central collisions at $\sqrt{s_{NN}}=130$ and 200 GeV and for Pb-Pb
  central collisions at $\sqrt{s_{NN}}=5.5$ TeV. Data taken from \cite{rhic}.
  The upper, central (solid
  lines) and lower limits of the theoretical uncertainty
  band correspond to ($Q_{0}\!=\!1$ 
  GeV, $\Delta Y\!=\!1$), ($Q_{0}\!=\!0.75$
  GeV, $\Delta Y\!=\!3$) and ($Q_{0}\!=\!1.25$ 
  GeV, $\Delta Y\!=\!0.5$) respectively, with $m\!=\!0.25$ GeV in all
  cases. 
}\label{fig3}
\end{figure}

With this set up we find a remarkably good agreement with the
pseudo-rapidity densities of charged particles measured in $0\!-\!6\%$
central Au+Au collisions at collision energies $\sqrt{s_{NN}}\iq130$
and 200 GeV.  The comparison with data \cite{rhic}, shown in \fig{fig2},
constrains the free parameters of the calculation to the ranges:
$Q_0\!\sim\!0.75\div1.25$ GeV, $m\!\sim\!0.25$ GeV and
$3\!\gtrsim\!\Delta Y_{ev}\!\gtrsim \!0.5$. These ranges determine the
uncertainty bands of the LHC extrapolation in \fig{fig3}.  The best
fits (solid lines in Figs. (\ref{fig2}) and (\ref{fig3})) are obtained
with $Q_0\iq1$ GeV, $m\iq0.25$ GeV and $\Delta Y_{ev}\iq1$. The
normalization constant, $C$, fixed at $\sqrt{s_{NN}}\iq130$ GeV and
$\eta\iq0$, is of order one in all cases.  The line of argument
that leads to these values is the following: First, the energy
extrapolation from $130$ to 200 GeV at central rapidities demands a
moderate evolution speed $\lambda\!\sim\!0.2$ \cite{hn}. From
\fig{fig1}, that condition is met by either initial saturation scales
$Q_0\sim 1$ GeV and small evolution rapidities $\Delta Y_{ev}\lesssim
3$ or at asymptotically large rapidities, $\Delta Y_{ev}\!\sim\!50$,
which are kinematically excluded. In the physically accessible range,
the solutions close to the scaling region, i.e. for $\Delta
Y_{ev}\!\sim\!10$, result in too narrow pseudo-rapidity distributions
independently of the value of $Q_0$, see \fig{fig2}B.  In the
pre-asymptotic regime at fixed $\Delta Y_{ev}\!\lesssim\!3$, those
solutions corresponding to a $Q_0\lesssim 0.75$ GeV yield exceedingly
broad distributions (see \fig{fig2}C).  Thus, the energy and the
pseudo-rapidity dependence independently constrain the parameters of
the gluon distributions probed at RHIC to the same ranges. This
provides the baseline for further evolution to LHC energies.  In
summary, these results indicate that the nuclear gluon densities
probed at RHIC are in the pre-asymptotic stage of the evolution.
This, together with the large values of the initial saturation scale
required by data suggests that the saturation of gold nuclei at RHIC
energies is not dynamically generated by the evolution but, most
likely, it is attributable to the nuclear enhancement factor that lies
at the basis of the MV model, i.e., to the fact that the number of
gluons in the nuclear wave function is large even at moderate energies
due to the spatial superposition of a large number of nucleon's gluon fields.

\begin{figure}[h]\epsfxsize=8.7cm
\centerline{\epsfbox{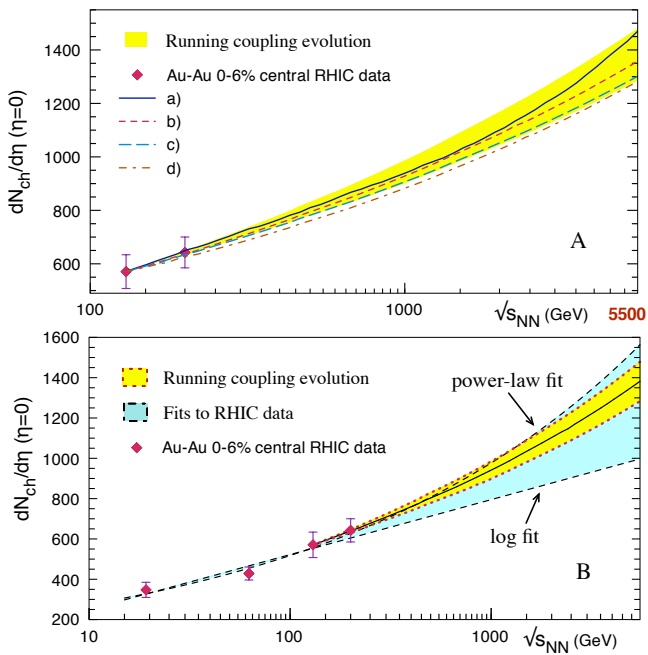}}
\caption{Charged particle multiplicity in central Au-Au collisions at
  $\eta=0$ versus collision energy.{\it Upper plot}: Results obtained
  with the set up leading to \eq{pred} (band) and several
  modifications of it (see text). 
  {\it Lower plot}: Power-law, $a\sqrt{s}^{\,b}$, and
  logarithmic, $a\!+\!b\ln s$, fits to RHIC data at
  $\sqrt{s_{NN}}=19.2$, 64.2, 130 and 200 GeV. 
}\label{fig4}
\end{figure}

The extrapolation to LHC energies, done neglecting the differences
between lead and gold nuclei and presented in \fig{fig3}, is now
straightforward and completely driven by the non-linear dynamics of
gluon densities. For central Pb-Pb collisions we get
\begin{equation}
  \label{pred}
 \left.\frac{dN^{Pb-Pb}_{ch}}{d^2b\,d\eta}(\sqrt{s_{NN}}\iq5.5\,\mbox{TeV},\eta=0)
\right.\sim1290\div1480\,,     
\end{equation}
with a central value corresponding to the best fits to RHIC data $\sim
1390$. These values are significantly smaller than those of other
saturation based calculations \cite{kln,asw,ekrt}, $\sim
1700\div2500$, and compatible with the ones based on studies of the
fragmentation region \cite{gsv}. Such reduction is due to the lower
speed of evolution yielded by \eq{frs} and to the proper treatment of
pre-asymptotic effects, thereby going beyond the scaling
ansatz. Importantly, the prediction for the midrapidity multiplicity
in \eq{pred} is very robust against changes in the description of
particle production and the implementation of large-$x$ effects. This
is illustrated in \fig{fig4}A, where the following modifications to
our set up have been considered (the itemization here follows the
labeling in \fig{fig4}A): a) Replacement of the ugd's in \eq{ktfact}
by the modified gluon distributions $h(Y,k)=k^2\, \nabla^2_k\,
\varphi(Y,k)$, as advocated in \cite{kt}. b) Regularization of the
strong coupling at the value $\alpha_{fr}\iq0.5$, c) Removal of the
$(1\!-\!x)^4$ corrections to the ugd's and d) Putting $m\iq0$. The
results obtained with these alternative configurations do not deviate
from the uncertainty band given in \eq{pred}, confirming that our
predictions are mostly driven by the properties of small-$x$
dynamics. Oppositely, our predictive power at large pseudo-rapidities,
$|\eta|\gtrsim 6$, is lessened by the sensitivity of the evolution to
the initial conditions and by our relatively crude implementation of
large-$x$ effects (see \fig{fig2}D), which are dominant in that
region.

Purely empirical parametrizations of
multiplicity data of a large variety of colliding systems allow a
logarithmic dependence on collision energy (see e.g. \cite{wb}).  As shown in 
\fig{fig4}B, RHIC data by themselves do not differentiate between this 
and other functional forms like power-laws, negating any possibility
to usefully constrain the 
expectations for LHC energies without further theoretical guidance.
Our results, similar to other calculations based on perturbative QCD, exhibit a
power-law growth of the midrapidity multiplicity with increasing
collision energy. The higher order corrections utilized here for the
first time provide
a richer physics input and result in a noticeably smaller power than previously
estimated. This fact is crucial to obtain a good description of both
the energy and pseudo-rapidity dependence of existing data and is the
key ingredient in the extrapolation to higher energies.
 
This research is sponsored in part by the U.S. Department of Energy
under Grant No. DE-FG02-05ER41377 and by an allocation of computing
time from the Ohio Supercomputer Center.
Useful discussions with Yuri Kovchegov, Anthony Kuhlman and
Heribert Weigert are gratefully acknowledged. 

 \end{document}